\documentclass[onecolumn]{revtex4}
\usepackage{epsfig}

\begin{document}
%%\runauthor{Rub\'{\i}, Gadomski}
%%\begin{frontmatter}

\title{Nonequilibrium thermodynamics versus model grain growth: derivation and some physical implications}  
\author{J.~M. Rub\'{\i}$^a$ and  A. Gadomski$^b$} 
%\author{ A. Gadomski$^b$ \corauth{}}
\affiliation{
$^a$Departament de F\'{\i}sica Fonamental, Universitat de Barcelona, \\
08028 Barcelona, Spain \\    
$^b$Institute of Mathematics and Physics, 
  University of Technology \\ and Agriculture,
85-796 Bydgoszcz, Poland 
}

%%\corauth[cor]{Corresponding author, fax:+48-52-340-8643, e-mail: 
%%agad@atr.bydgoszcz.pl}

%\date{}

%\newpage
\begin{abstract}
% \baselineskip8mm
%\noindent 
Nonequilibrium thermodynamic formalism is proposed to derive the flux of grainy (bubbles-containing) matter, emerging in nucleation and growth process. Some power- and non-power limits, due to the applied potential as well as owing to basic correlations in such systems, have been discussed. Some encouragement for such a discussion comes from the fact that the nucleation and growth processes studied, and their kinetics, are frequently reported in literature as self-similar (characteristic of algebraic correlations and laws) both in basic entity (grain; bubble) size as well as time scales.

\end{abstract}

%\newpage
% \baselineskip9mm
%\begin{PACS}
% PACS numbers: 05.40.-a, 64.60.-i, 81.10.Jt, 82.70.Rr
%\end{PACS}
%\begin{keyword}
%nonequilibrium thermodynamics, Onsager coefficients, Gibbs equation, entropic barrier, curvature, 
%nucleation-and-growth phenomena, Fokker-Planck equation,  power laws, random walk, thermodynamic and physical potentials \\
%{\PACS 64.60.-i, 05.40.-a, 81.10.Jt, 82.70.Rr}
%\end{keyword}
%\end{frontmatter}
\maketitle

%%%%%%%%%%%%%%%%%%%%%%%%%%%%%%%%%%%%%%%%%%%%%%%%%%%%%%%%%%%%%%%%%%%%%%%%%%%%
\section{Introduction}

Nucleation and growth (NG) phenomena appear to be ubiquitous in materials processing, nanotechnology, metallurgy and biotechnology. Without loosing generality they can verbally be taken under common denominator termed most frequently recrystallization and grain growth though such an overall notion implicitly assumes that some ordering is present in the system \cite{shvind}. Certainly, if the underlying ordered physical microstructure is not present, one may quite generally think about bubbles-containing systems, like soap froths, foams or certain sponge-resembling assemblies \cite{hweaire}. 

The main signatures of such phenomena as model systems are as follows \cite{mweaire}: \\
(i) They  take mostly place under certain constraints concerning the constancy or non-constancy of the overall system volume; 
(ii) The process relies on evolution of grains constituting a specimen in such a way that the mean radius of the grains (bubbles) has to grow in time whereas their total number has to decrease;  
(iii) An observed physical tendency while going over sufficient number of time intervals is a drop of the grain (bubble) surface free energy in such a way that a low surface free energy microstructure is eventually reached. 

As expected, there are very many quite important details that can be taken into account to provide a successful description of NG-processes, especially of their kinetics. They have thoroughly been presented in \cite{novikov}, chap. 3. They either emphasize a pronounced role of topology, dealing with such NG-systems as composed of convex polyhedra, or they offer a similar metric and combined approach, in two-dimensional systems represented mostly by Mullins--Von Neumann evolution formula valid for mean area of $n$-sided grains \cite{mullins}. (A Monte Carlo model for three-dimensional case has also been proposed recently \cite{liu}.) 
The latter is also proposed to split into three groups: (i) NG phenomenon, here of the so-called
normal type \cite{mullins,novikov}, as a deterministic process governed by a continuity equation; (ii) NG of normal type as standard diffusion along grain size axis; (iii) NG of normal type as a stochastic process, considered in terms of grain (bubble) boundary motion, with a Gaussian noisy source, exclusively, \cite{pande} and refs. therein. 

While listed above the basic signatures of NG-processes we have right now to add an important missing point. Namely, that at least within a statistical-mechanical framework, they are most successfully considered when a local conservation law is applied. It is mostly so on the so-called mesoscopic level of description, in which the system is in principle treated as macroscopic but where some prevalence of quantum-oriented effects, {\it viz} correlations, appears. No doubt that such a description leads to the well-known Fokker--Planck--Kolmogorov (FPK) formalism, with some possibilities of getting useful solutions, mostly after applying suitable, physically justified, boundary conditions and comfortably assuming the phase space to be semi-infinite. (Some visible difficulties in providing the solutions may emerge when a finite phase space is taken, an assumption provoked mostly by experimental situations \cite{agluczka}.) Moreover, such a mescoscopic description has proved to be useful in nanoscale, that means, in a physical scale in which emergence of new materials is nowadays very desirable because of possible applications in medicine, microelectronics, pharmacy to mention but a few. 

A crucial point, however, in proposing a FPK-type description, with suitable initial and boundary conditions (IBCs), is to construct the flux in the space of grain (bubble) sizes. 

In a series of papers \cite{niemiec,agad1,agad2}, the corresponding matter flux has been contructed in a phenomenological way, in which to a standard diffusion (1st Fick's law) term a drift curvature - (capillarity) dependent term has arbitrarily been added just for reflecting properly the essential physics staying behind it \cite{novikov,mullins,pande}. 

Recently, another more fundamental way of constructing the flux has been offered. It is based on dealing with a NG-system as an open thermodynamic system, and on applying the principles of nonequilibrium thermodynamics in the space of grain (bubbles) sizes. This immediately leads to express the entropy production by the above mentioned flux as well as by the corresponding chemical potential gradient. The gradient, in turn, has to be defined by the chemical activity (not equal to one for non-ideal systems) and by the presumed, but physically very accepted under concrete circumstances, physical potential that can be chosen being either of power- or of non-power form in the grain (bubble) size. The procedure just described can be found elsewhere \cite{rubi1,rubi2,rubi3}. 

The paper's contents is distributed as follows. In Section 2, we sketch the  
Louat--Mulheran--Harding  (LMH) model describing a NG-process of uniform boudnaries (or, of normal type \cite{mullins}).  
In Sections 3 and 4, respectively, we propose a derivation of the flux of grainy (bubbles-containing)
matter based on the nonequilibrium thermodynamics formalism (Sec. 3) whereas in the next section (Sec.
4) we discuss some important implications that may be considered in the new proposed physical context near thermodynamic equilibrium, mostly. 
Section 5 serves for concluding remarks preferentially within the framework of NG polycrystalline
evolutions \cite{nov,shvind}.

\section{ Louat--Mulheran--Harding  (LMH) model: A sketch}

\indent    The LMH model \cite{mulheranh} is a bit aside or slightly forgotten statistical-mechanical model that was originally invented by Louat for describing evolutions of polycrystalline NG-agglomerates in dimension one \cite{louat}. Inspite prescribed boundary conditions, it was thought of to be equivalent to the standard Einstein Random Walk (RW) though realized not in a position space but in the space of grain (crystallite) sizes. As in the precursor model, the diffusion coefficient was assumed to be a constant since an incorporation of the grain surface magnitude, proposed later by Mulheran 
and Harding in higher dimensions, is certainly not possible in the one-dimensional space. 

Although the limitations of those statistical-mechanical (ordered)  agglomeration models are obvious presently (the specialists often argue that the LMH-model becomes partly useful while describing some evolutions in nanoscale what seems likely to be its biggest advantage), it is still worth facing its main ideas. 

The clue idea is that the original LMH-model \cite{louat,mulheranh} is a conservative model in which a flux of grainy matter is supposed to be of the form
\begin{eqnarray}
\label{J}
J(v, t) = - \sigma v^{\alpha -1} f(v,t)
 - D v^{\alpha} {\partial \over \partial v}  f(v,t), 
\end{eqnarray}
which means that it is decomposed into two parts. Here $\sigma $ and $D$ are surface tension as well as diffusion reference constants, respectively, and generally 
$\sigma \ne D$  holds; the independent variables $v$ and $t$ will be defined below. $\alpha $ stands for a dimensionality $d$ (where $d=1,2,3, ...$) dependent quantity, namely 
\begin{eqnarray}   \label{alfa} 
\alpha = 1-\frac{1}{d}.
\end{eqnarray}
The first part of the right-hand side (r.h.s.) of Eq. (\ref{J}) is a drift term, whereas the second part of the r.h.s. of Eq. (\ref{J}) stands for a diffusion term. The former is a capillarity term, proportional to the grain curvature, and strictly related with the well-known Laplace-Kelvin-Young law, so much explored for bubbles-containing systems \cite{hweaire,agluczka,agad3} in which a pressure difference appears to be a main driving force. 
The latter is simply the first Fick's law, and emphasizes a proportionality of the flux to the grain surface magnitude. (Such a term, as it stands, may constitute the overall flux alone: This observation has extensively been explored to describe certain agglomerations under negligible surface or line tension regime, {\it i.e.} explicitly without the curvature-containing drift term. The system becomes, however, unconstrained as far as its total volume is concerned \cite{mniemiec}.) Both the terms incorporated in (\ref{J}) are based on phenomenological laws, so is also the construction of the mixed convection-diffusion flux (\ref{J}). 

After constructing the flux, we have to plug it in the well-known continuity  equation   
\begin{eqnarray}
\label{Eq}
\frac{\partial}{\partial t} f(v,t) =  
- {\partial \over \partial v} J(v, t), 
\end{eqnarray}
where $v$ is the volume of a grain, $f(v,t)$ is the distribution function of the grains
 at time $t$ (being of the meaning of the number density), that means, $f(v,t)dv$ is a relative number  of grains of size in the volume range $[v,v+dv]$. 

The overall conceptual and mathematical construction is to be completed by the ICS. 
The suitable IC looks like 
\begin{eqnarray}       \label{t=0}
f(v,t=0) =  F_0(v),
\end{eqnarray}
where $F_0(v)$ is a given initial distribution of grains; in many studies \cite{mulheranh,louat,agad2,agad3} $F_0(v)$  was assumed to be of delta--Dirac form but some other realizations are also possible \cite{niemiec}. A general conclusion is, however, that the process in question does not depend upon an IC, {\it cf.} \cite{niemiec}, and refs. therein. After a sufficient time spane being overcome it completely forgets its initial state which is a case quite often met in many open thermodynamic systems. 

This is not the case of the BCs, in turn. The process substantially depends upon BCs chosen. 
A choice that remains as most explored is the choice of zero Dirichlet BCs, namely 
\begin{eqnarray}       \label{bound0}
f(v=0,t) = f(v= \infty,t) =  0. 
\end{eqnarray}
(Notice that the LMH-model does work in a semi-infinite phase space, $v\in[0,\infty ]$; it may lead to  some inconsistency, because volume of the  
individual grain can be arbitrary large and can even be distinctly larger than 
 the overall volume of the system as a whole, {\it cf.} discussion in \cite{agluczka,agad3}.) 

There is some physics staying behind the BC: The grains of zero as well as of infinite size have zero account for the process studied. This physical constraint is sometimes called a normality condition, and the NG-process is supposed to be normal when the above is true. Otherwise, one may term the agglomeration abnormal. 

For the LMH model, which is formally the case of $\sigma = \alpha$ in Eq. (\ref{J}) the average total volume of the whole system remains constant in  
time and takes on a finite value. This is another constraint of the LMH-agglomeration, namely, that it takes place under constant total volume condition. This is, by the way, not the case of other evolutions with negligible surface tension effect, where capillary forces do not drive the evolving assembly \cite{mniemiec}. In such a system, as is ascertained above, we have to get rid of the curvature dependent first term of the r.h.s. of Eq. (\ref{J}). 
Moreover, realize that we practically may discard the curvature-dependent term if we enter
sufficiently mature growth stage, presumed that the grain radii take then on quite large values
so that their curvatures become very small, {\it cf.} \cite{niemiec,mniemiec}. 

What remains to be done when the problem is formulated (see the above equations) is just to solve
it. It was recently done by means of variables' separation method not only for original
LMH-system but also for its physically interesting variations (modifications) both in finie as
well as infinite phase spaces, and can be found elsewhere 
\cite{mulheranh,agad1,agad2,agad3,niemiec}. Some important physical quantities of the evolving
agglomerate have been calculated based on the first three statistical moments of the process.
Then zeroth moment has always been related to the average number of grains in the system (it
mostly suffers from an algebraic drop with time for $t>>1$), and the first to its overall volume
which is a constant value. From them the mean grain radius can be obtained
\cite{agad2,agad1,mulheranh}: It behaves powerly with time $t$ in an asymptotic time limit, and
the growth exponent is $1/(d+1)$. These compose the main characteristics of the model LMH-process
just described. (From the second and first moments a useful information about volume fluctuations
can additionaly be taken \cite{mniemiec,agad2}.)

In the subsequent section, we will be still interested in a construction procedure of the flux $J(v,t)$, hopefully arriving at a form resembling that of Eq. (\ref{J}). But another clearly non-phenomenological method will be applied therein though we will be not capable of avoiding phenomenology completely, see the discussion concerning the physical potential $\phi$ below. We will, however, be able of speculate firmly and in a physically reasonable way on the basic mechanism of the agglomeration in the spirit of incorporating (or, having not incorporated) the suitable entropic barrier as geometrical one \cite{rubipre}, here inevitably associated with the grain curvature. 

\section{Nonequilibrium thermodynamics derivation of LMH--like grain growth model}

Assume as above that the continuity equation (\ref{Eq}) is the evolution equation for the
agglomerating system but the flux $J(v,t)$ remains unspecified. Assume, additionaly, that there is a method of constructing it. The method is called mesoscopic nonequilibrium thermodynamics \cite{rubi1,rubi2,rubi3}, and starts with the Gibbs equation which represents the entropy virtual change, namely \cite{rubi1,rubi3}
\begin{eqnarray}       \label{Gibbs}
\delta S = - {1\over T} {\int \mu (v,t) \delta f \;dv} ,
\end{eqnarray}
where $f\equiv f(v,t)$, $T$ is the temperature, and $\mu (v,t)$ is the generally time-dependent chemical potential in $v$-space. The latter is given by ($\mu \equiv \mu (v,t)$ for brevity is taken)
\begin{eqnarray}       \label{cpot}
\mu = {k_B} T\quad ln (a f) ,
\end{eqnarray}
where $a\equiv a(v)$ is an activity coefficient expressing the fact that the system is not ideal; for
$a=1$ it would be ideal. (Notice that the $ln$-form of the chemical potential suggests the system were ideal but,  fortunately, the condition $a\ne 1$ also violates this ideality.) $k_B$ is the Boltzmann constant. Next, $a$ is given in terms of a potential that for the reason of differentiation against the chemical potential $\mu $ let us call the physical potential, $\phi $. $a$ reads now \cite{rubi1,rubi3}
\begin{eqnarray}       \label{ppot}
a = e^{\phi/{k_B T}}.
\end{eqnarray}
Taking the temporal derivative in Eq. (\ref{Gibbs}) and performing a partial integration (assuming, however, that $J\equiv J(v,t)$ vanishes at the ends of the phase space), one arrives at the entropy production, $\sigma $, in $v$-space
\begin{eqnarray}       \label{entropy}
\sigma = - {1\over T} J {{\partial \mu}\over  {\partial v}} 
\end{eqnarray}
from which we infer the expression for the current 
\begin{eqnarray}       \label{current}
J = - {1\over T} L(v) {{\partial \mu}\over  {\partial v}}. 
\end{eqnarray}
We have assumed that the process is local in $v$. In view of Eqs (\ref{current}), (\ref{cpot}) and (\ref{ppot}) we obtain
\begin{eqnarray}       \label{current1}
J = - {1\over {T f}} L(v) \bigg[{{k_B T}{\partial f\over  \partial v}+ f {\partial \phi\over  \partial v}}\bigg]. 
\end{eqnarray}
Let us define the mobility $b(v)$ as $b(v) = {1\over {T f}} L(v) = {D\over {k_B T}} {v^\alpha }$, where $D$ and $\alpha $ may have their role played in accordance with what has been stated in the preceding section, where they have been specified; this will somehow suggest us the mechanism for the entropic barrier mentioned above (a trial toward self-similarity will also be considered), and its connection with the physical potential $\phi $. 
The obtained flux $J$ reads finally
\begin{eqnarray}
\label{JJ}
J = - D v^{\alpha} {\partial \over \partial v}  f  - b(v) f {\partial \over \partial v}  \phi .
\end{eqnarray}
The expression derived, Eq. (\ref{JJ}), is quite general (inspite of the power-law form assumed in the Onsager coefficient $L(v)$). The main physics to play now is to propose a valuable expresion for the physical potential $\phi = \phi (v)$. This is, however, a matter of proposing a basic agglomeration mechanism for the system to evolve. It will be thoroughly discussed in the next section by considering power- as well as non-power forms in $v$. 

\section{Physical implications: Power and non-power laws applied} 

In \cite{rubi1} it was shown that diffusion as well as mobility functions are very related to each other. They are also related with the Onsager coefficient $L(v)$ via the density function $f$, or also via an inclusion of the temperature $T$ as is in the case of mobility, and does not apply for the diffusion coefficient. 

The diffusion coefficient (in its full form denoted by $D(v,t)$) is generally assumed to be defined by the Green-Kubo correlation formula, in which, as integrands, the random parts, $J^r (v,t)$, of the flux (\ref{JJ}) of grainy matter, in two suitable time instants, are involved. It reads 
\begin{eqnarray}       \label{Onsager}
D(v,t) = {{k_B}\over f} L(v) = D {v^\alpha }, 
\end{eqnarray}
that means that algebraic correlations are assumed in the correlator based on $J^r (v,t)$ \cite{rubi2,rubi3}. 

Since it has been argued \cite{mullins} (see, literature therein), and experimentally proved \cite{kurz}, that the NG-processes, namely recrystallization as well as grain growth are self-similar in many respects and when based on many versatile measures of self-similarity both in space (grain size) and time, it would be useful to test this assertion. 

Our first proposal is to assume, inspite of the algebraic (self-similar-in-time) correlations of the flux $J^r (v,t)$ also that the 
physical potential is of a power form 
\begin{eqnarray}       \label{fi-power}
\phi=  {{\phi _o}\over {v^\epsilon}}, 
\end{eqnarray}
where $\phi _o$ is a constant (it may depend upon temperature, and as suggests the construction of the Lennard-Jones (LJ) potential, it should be nonpositive at least, or negative even; note that the attractive part of the LJ-potential has to be taken with minus sign, and we need the Van der Waals-like attraction for preserving agglomeration), and $\epsilon $ is a positive exponent. 

We have now to put  ${{\partial \over \partial v}  \phi } = - {{\phi_ o \epsilon}\over {v^{\epsilon +1}}}$ into Eq. (\ref{JJ}), and the result is as follows
\begin{eqnarray}
\label{JJp}
J = - D v^{\alpha} {\partial \over \partial v}  f  - \sigma {v^\chi} f ,
\end{eqnarray}
where $\chi = \alpha - \epsilon -1$ and $\sigma = - {\phi _o} \epsilon {D\over {k_B T}}$. This way, we get the grainy flux (\ref{J}), or strictly speaking (\ref{JJ}), in a very satisfactory form. A direct comparison with (\ref{J}) from Sec. 2 suggests to let $\epsilon \to 0$ ($\chi \to \alpha - 1$) which yields $\phi \to const$ (very weak forces as the Van der Waals typically are!). Thus, the flux (\ref{JJp}) is really in the form we await very much except that the very fact that the surface tension reference constant $\sigma $  unfortunately goes to zero, see above. This way, the drift term is completely washed out, and we get no drift effect on the agglomeration. Thus, an entropic barrier due to curvature (bear in mind that  $v^{\alpha -1} \propto {1\over R}$, where $R$ is a grain radius, and when $\alpha $ is taken from Eq. (\ref{alfa})) cannot be proposed based on the power-like form of the physical potential $\phi $. This is, qualitatively speaking, the case considered in \cite{mniemiec} and mentioned before. 

In contrast to the above, there is a possibility of including such a geometrical barrier when a
non-power (here, logarithmic) form of the potential is presumed. The way of doing that is worth
exploring since no one can exaggerate the role of curvature in physics, especially in
agglomeration processes \cite{bedo,agad3}. The first correction in curvature, {\it e.g.} due to the surface tension and named usually twice the mean curvature, is realized to be the main driving force (the whole thermodynamic context assumes the Gibbs-Thomson or capillary length to be a  characteristic length of surface tension shrinking action) of some oversimplified grain growth mechanism \cite{hillert} whereas the so-called second correction, frequently related with the Tolman length \cite{tolman,bedo,agad3} and named the Gaussian curvature, shows up certain elastic (bending, and insistance to bending termed often rigidity) properties of an agglomerate in a time instant chosen, and when going over its evolution stages, in which such characteristics can certainly change. 

Thus, the proposed non-power form of the physical potential is of a logarithmic form, namely 
\begin{eqnarray}       \label{fi-npower}
\phi= {\phi _o}\quad {ln(v/{v_o})},    
\end{eqnarray}
where $v_o >0$ stands for the initial grain volume. Amazingly, exactly this form suits very well to what we want to get. The entire rationale we try to reveal thoroughly throughout the present section is that we wish to have in the drift term, present in either Eq. (\ref{J}) or Eq. (\ref{JJ}), a factor of $v^{\alpha -1}$ which is explicitely a clear signature of the curvature, being a reciprocal of the grain radius $R$. Thus, after simple algebra 
${{\partial \over \partial v}  \phi } = {{{\phi _o}\over {v}}}$, {\it mutatis mutandis}, we are able
to recover the desired form of the drift term in flux (\ref{J}), and this way both the fluxes,
(\ref{J}) and (\ref{JJ}) take firmly on the same form, presumed that $\alpha $ is that dimensionality-
dependent fractional number taken from (\ref{alfa}). To have a really entropic barrier an additional
presumption on explicit dependence of the prefactor ${\phi _o}$ upon the thermal energy ${k_B}T$ is
needed, as applied for star polymer complex solutions, for example \cite{waclaw}, carrying however
less about that their logarithmic part of the potential is distance-dependent, whereas ours
explicitely not, {\it cf.} Eq. (\ref{fi-npower}). (We may have some excuse here by realizing that $v\propto R^d$, where d is the space dimension, and $R$ can likely better mimic a distance, or implicitly a position measure than $v$ does.) 

Another issue that can be addressed is of more fundamental nature. It concerns our analogy between the RW in a position space and the walk along the grain size axis, see beginning of Sec. 2. There are some experimental evidences \cite{kurtz} as well as theoretical predictions \cite{pande,agad3} (look at refs. therein) that the walk in grain-size space would belong to a broad class of geometric Brownian motion, the probability distribution of which is a logarithmic Gaussian \cite{paulb}. Thus, instead of $v$ in the mathematical form of Gaussian distribution a $ln(v)$ must appear. Therefore such a potential seems legitimate too. 

Accepting that the key problem, because of the RW-analogy in both the spaces mentioned, does not
totally rely on  whether we have $v$ or $R$ in potential function argument, {\it i.e.} a distance
measure in (\ref{fi-npower}), we may at least for a two-dimensional case, invoke another experimental
study with a logarithmic potential. The study is on a confined mesoscopic system in which the
so-called Wigner islands appear. These are charged balls interacting via an electrostatic potential
\cite{wigner}. They represent the vortices in type-I superconducting systems. It turns out that just
the logarithmic interaction potential causes maximum compaction of the charged balls of milimeter
(macroscopic) size, typically moving and agglomerating on a conductor plane \cite{wigner}. This is
then probably no surprise that the $ln$-potential (\ref{fi-npower}) but not that power-like (\ref{fi-power}) gives more chances for the system to evolve in a constant total volume regime as the above mentioned LMH-system does \cite{mulheranh,novikov,agluczka} but another one does not \cite{mniemiec}. This is, as is stated in Sec. 2, the case when the governing  mechanism based on curvature driving dominates, and in which a surface (line) tension effect cannot be avoided. 

\section{Concluding remarks}

\indent  The concluding remarks are juxtaposed in the following way: \\
\begin{enumerate}
\item We have applied the nonequilibrium mesoscopic thermodynamics in order to justify a phenomenological construction of the grainy matter flux given in Sec. 3 by Eq. (\ref{JJ}) whereas in Sec. 2 by Eq. (\ref{J}). The fundamental procedure offered by nonequilibrium theromodynamics method stresses mostly the role of correlations or grainy matter fluctuations in the random part of the flux, as well as emphasizes a choice of the physical potential applied \cite{rubi1,rubi2,rubi3}. Since we know that, inspite of many additional sub-mechanisms manifesting in the course of agglomeration (misorientation effects; pinning of grain boundaries; dislocation density gradients' action, segregation of molecules, and alike \cite{novikov,shvind}), capillary forces are those governing the system, and we may be glad because of finding a "compaction" $ln$-potential which recovers perfectly the basic form of the flux (\ref{J}). Such a proposal is to our knowledge original and new. (It is worth noticing that the compaction effect, mostly forcing the system to exist under constant volume circumstances, where however the overall grains' evolution scenario proceeds in a random manner, would quite closely resemble the close-packing effect so well pronounced in amorphous as well as polycrystalline materials under evolution \cite{agluczka}. A signature thereof is realized to be the value of the growth exponent which is $1/(d+1)$, as is mentioned in Sec. 2, and what is characteristic of the LMH-like systems \cite{mulheranh,agad2,agad3,niemiec}.)
This could be anticipated as another success of the theory that before described quite exhaustively homogeneous
\cite{rubi1,rubi3} as well as heterogeneous complex nucleations \cite{rubi2}, se also Fig. 1 for sketch of the
conception of random close-packing.
\item As was stated in Sec. 2, the LMH-system always evolves under a constant volume condition,
therefore a pivotal role of surface tension in it is absolutely no surprise (the parameter $\sigma $ in
(\ref{J}) or in (\ref{JJ}) must clearly be nonzero). Therefore also the mentioned involvement of the curvature
term is needed just in order to have a visible  nonzero effect in the drift term what ist quite important when
$t >> 1$, that means, for a mature stage of evolution when the grain radii are typically big but curvatures are
rather small.  
\item Applying the nonequilibrium thermodynamics  method to a relatively simple and soluble
\cite{agad1,agad2,agluczka} LMH- or alike \cite{agad2,agad3,pande,mullins} systems opens up new
possibilities to understand that the geometrical-physical "object", which the curvature is, may very
likely stand as a solely picked up entropic barrier, though a task remains to find the form of the prefactor in (\ref{fi-npower}) in exact form and being temperature-dependent \cite{rubipre,waclaw}. 
\item The forms of the proposed physical potentials are offered neither in unique nor in a systematic ways: A choice of potentials known in physics is very broad \cite{stauffer}. But we have proposed two of them which reveals  some physics of the NG-processes with non-negligible (LMH) as well as negligible (interacting clusters formation \cite{mniemiec}) surface tension effects. 
\item The RW-analogy emphasized so much could be an advantage; it may, however, yield also some 'uncertaintes', questionmarks, as for example the exact form of the physical potential and how to go out from the position or distance space to the grain volume or size space. Another drawback appears to be that even the form of the evolution equation (\ref{Eq}), no matter whether with or without the curvature-dependent term, is not unique: Van Kampen, for example, proposes a third form, in which the diffusion function is placed in front of the Laplace operator on the r.h.s of Eq. (\ref{Eq}) \cite{kampen}. Therefore, generally speaking, the local conservation law may also be questioned. 
\item Except the agglomeration in polycrystalline model materials very often mentioned throughout the paper,
one can expect a certain appreciable application of the offered modeling in another phenomenon, termed 
dropwise coalescence, in which the Laplace-Kelvin-Young law (see, Sec. 2), containing the capillarity-dependent
drift term but with another $\sigma$, modified by the ratio of the satured vapor as well as liquid densities,
respectively, is to be taken into account to describe readily the droplets' agglomeration \cite{landau},
presumably with constantly increasing agglomerate's total volume \cite{mniemiec}.
\end{enumerate}

%\footnote{}

\section*{Acknowledgement}
{\normalsize One of us (A.G.) would like to take very much the opportunity of thanking Dr. Andrej
Jamnik from the University of Ljubljana, Slovenia,  for reminding a useful notion of the {\it random
close-packing} ( rcp) phase in condensed-matter physics which helps to understand many properties of
random matter agglomerations. J. M. R. wants to thank Dr. David Reguera for fruitful discussions.}

% \newpage

\newpage

\begin{figure}
\begin{center}
\includegraphics[scale=.5]{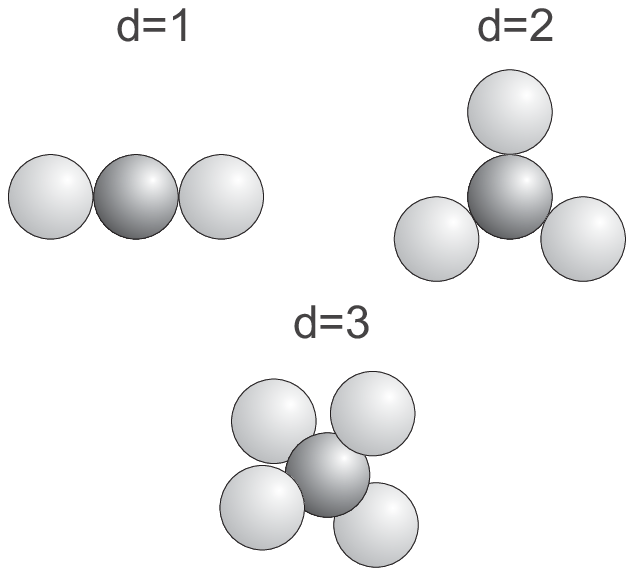}
\caption{{Random close-packing or matter compaction effect realized in systems evolving under a
constraint of constant total volume. The nearest (and non-intersecting) neighborhood of a shadowed middle grain
chosen is composed always of $d+1$ nearest neighbors in a $d$-dimensional Euclidean space; here, a demo on the
spheroidal grains is proposed for simplicity.
}}
\end{center}
\end{figure}
%\noindent {\bf Figure Caption}

%\noindent {\bf Figure 1.} Random close-packing or matter compaction effect realized in systems evolving under a
%constraint of constant total volume. The nearest (and non-intersecting) neighborhood of a shadowed middle grain
%chosen is composed always of $d+1$ nearest neighbors in a $d$-dimensional Euclidean space; here, a demo on the
%spheroidal grains is proposed for simplicity.


\begin{thebibliography}{10}
\bibitem{shvind} G. Gottstein, D.A. Molodov  (Eds.), Recrystallization and Grain Growth. Proceedings of the First Joint International Conference on Recrystallization and Grain Growth, Springer--Verlag, Berlin, 2001.
%\bibitem{thompson} J. Cho, C.V. Thompson, Appl. Phys. Lett. 54 (1989) 2577. 
%\bibitem{peczak} P. Peczak, G.S. Grest, D. Levine, Phys. Rev. E 48, 4470 (1993) 
\bibitem{hweaire} D. Weaire, S. Hutzler, The Physics of Foams, Oxford University Press, Oxford, 2000. 
\bibitem{mweaire} D. Weaire, S. McMurry, Solid State Phys. 50 (1997) 1.
\bibitem{novikov} V. Novikov, Grain Growth and Control of
Microstructure
and Texture in Polycrystalline Materials, CRC Press, Boca Raton, 1997.
\bibitem{mullins} W.W. Mullins, Acta mater. 46 (1998) 6219.
\bibitem{liu} G. Liu, H. Yu, X. Song, X. Qin, Materials \& Design 22 (2001) 33. 
\bibitem{pande} C.S. Pand\'e, A.K. Rajagopal, Acta mater. 50 (2002) 3013.
\bibitem{agluczka} A. Gadomski, J. \L uczka, R. Rudnicki, Physica A, {\it to be published}. 
\bibitem{niemiec} M. Niemiec, A. Gadomski, J. \L uczka, Acta Phys. Polon. B 32 (2001) 581; 1513.
\bibitem{agad1} A. Gadomski, Physica A 274 (1999) 325.  
\bibitem{agad2} A. Gadomski,  Nonlinear Phenomena in Complex Systems 3 (2000) 321. 
\bibitem{rubi1} D.Reguera, J.M. Rub\'{\i}, A. P\'{e}rez-Madrid, Physica A 259 (1998) 10.
\bibitem{rubi2} D.Reguera, J.M. Rub\'{\i}, L.L. Bonilla, {\it in}: Mathematical Modelling for Polymer Processing, V. Capasso (Ed.), Springer--Verlag, Berlin, 2002, vol. 2, chap. 3 (Mathematics in Industry Series). 
\bibitem{rubi3} D.Reguera, J.M. Rub\'{\i}, A. P\'{e}rez-Madrid, J. Chem. Phys. 109  (1998) 5987. 
\bibitem{nov} S.P. Marsh, C.S. Pand\'e (Eds.), Modeling of 
Coarsening and Grain Growth, TMS, Chicago, 1993. 
\bibitem{mulheranh} P.A. Mulheran Acta metall.mater. 40 (1992) 1827.
\bibitem{louat} N.P. Louat, Acta metall. 22 (1974) 721.
\bibitem{agad3} A. Gadomski, J. \L uczka, Acta Phys. Polon. B 33 (2002) 1131. 
\bibitem{mniemiec} M. Niemiec, A. Gadomski, J. \L uczka, L. Schimansky--Geier, Physica A 248 (1998) 365. 
\bibitem{rubipre} D. Reguera, J.M. Rub\'{\i}, Phys. Rev. E 64 (2001) 061106.
\bibitem{kurz} K.J. Kurzyd\l owski, B. Ralph, The Quantitative Description of the Microstructure of Materials, 
CRC Press, Boca Raton, 1997.
\bibitem{bedo} E.M. Blokhuis, D. Bedeaux, HCR Advanced-Education Review, John Wiley \& Sons, New York, 1994, pp. 55--68. 
\bibitem{hillert} M. Hillert, Acta metall. 13 (1965) 227. 
\bibitem{tolman} R.C. Tolman, J. Chem. Phys. 17 (1949) 118.
\bibitem{waclaw} M. Watzlawek, C.N. Likos, H. L\"owen, Phys. Rev. Lett. 82 (1999) 5289. 
\bibitem{kurtz} S.K. Kurtz, F.M.A. Carpay, J. Appl. Phys. 51 (1980) 5125; 5745. 
\bibitem{paulb} W. Paul, J. Baschnagel, Stochastic Processes. From Physics to Finance, Springer, Berlin, 1999. 
\bibitem{wigner} M. Saint Jean, C. Even, C. Guthmann, Europhys. Lett. 55 (2001) 45. 
\bibitem{stauffer} D. Stauffer (Ed.), Annual Reviews of Computational Physics IX, World Scientific, Singapore, 2001, pp. 1--104. 
\bibitem{kampen} N. Van Kampen, Z. Phys. B 68 (1987) 135. 
\bibitem{landau} L. Landau, A. I. Akhiyezer, E. M. Lifshitz, Course of General Physics. Mechanics and Molecular
Physics, Nauka, Moscow, 1965, chap. 12 (in Russian). 
\end{thebibliography}
\end{document}